\documentclass{elsart5p}
\usepackage{cite}
\usepackage{amsmath}
\usepackage{amssymb}
\usepackage{bm}

\newcommand{\im}{\mathrm{i}}
\newcommand{\Rt}{{\widetilde R}}
\newcommand{\rhot}{{\tilde \rho}}
\newcommand{\Lt}{{\tilde{\mathcal{L}}}}
\newcommand{\diff}{\mathrm{d}}
\newcommand{\version}{\date}
\DeclareMathOperator{\trace}{tr}

\DeclareMathOperator{\Imag}{Im}

\begin{document}
\begin{frontmatter}
\title{On the conundrum  of deriving  exact solutions\\from
approximate master equations}

\author{Roland Doll},
\author{David Zueco},
\author{Martijn Wubs},
\author{Sigmund Kohler},
\author{Peter H\"anggi}

\address{Institut f{\"u}r Physik, Universit{\"a}t Augsburg,
Universit\"atsstra{\ss}e 1, D-86135 Augsburg, Germany}

\date{\version}

\begin{abstract}
We derive the exact time-evolution for a general quantum system under
the influence of pure phase noise and demonstrate that for a Gaussian initial
state of the bath, the exact result can be obtained also within a perturbative
time-local master equation approach already in second order of the system-bath
coupling strength. We reveal that this equivalence holds if the initial state of
the bath can be mapped to a Gaussian phase-space distribution function.
Moreover, we discuss the relation to the standard Bloch-Redfield approach.
\end{abstract}


\end{frontmatter}

\section{Introduction}

The coherent evolution of small quantum systems is typically influenced by its
interaction with environmental degrees of freedom, which results in quantum
dissipation and decoherence.  These ubiquitous phenomena play a crucial role in
various fields of physics and chemistry, such as quantum optics, electron
transfer reactions \cite{Kuhl2000, Egorova2003}, the electron transport through
molecular wires \cite{Cizek2004a,Kohler2005a}, and in particular in quantum information processing, where we
recently witnessed significant experimental progress \cite{Vion 2002a,
Chiorescu2003a, Yamamoto2003a}.  The optimisation of the coherence properties of
quantum devices certainly requires a good theoretical understanding of the
processes that induce decoherence.

The environment of a quantum system is frequently modelled as an
ensemble of harmonic oscillators that couple to the system
\cite{Magalinskii1959a, Caldeira1983a, Leggett1987a, Hanggi1990a}. If
the coupling is linear in the oscillator position, one can formally
eliminate the environment to obtain a closed equation for the
dissipative quantum system. Such equations are in general not easy to
deal with and, accordingly, only a few exact solutions exist in
dissipative quantum mechanics, e.g.\ for the dissipative harmonic
oscillator \cite{Grabert1984a, Grabert1984b, Grabert1988a,
Riseborough1985a} and its parametrically driven version
\cite{Zerbe1995a}. Recently, an exact solution has been found also for
the dissipative Landau-Zener problem at zero temperature
\cite{Wubs2006a}.  A whole class of system-bath models that can be
solved exactly are those in which the system Hamiltonian and the
system-bath coupling commute \cite{Luczka1990a, vanKampen1995a,
Palma1996a, Duan1998a, Mozyrsky1998a, Braun2001a, Krummheuer2002a,
Doll2006a, Doll2007b, Solenov2007a}. Herein we focus on such so-called
pure phase-noise models.

Even though pure phase noise allows an exact solution of the reduced
quantum dynamics, it is sometimes convenient to employ a perturbative
master equation approach, such as the Bloch-Redfield equation
\cite{Blum1996a}. On the one hand, those approaches may provide direct
access to the dephasing rates avoiding tedious algebra and in
particular in the limit of weak system-bath coupling, they are
expected to give quantitatively good results. On the other hand, it is
possible to test their quality when an exact solution is available. We
here show that for phase noise, the results of approximate master
equations can even be exact, despite the fact that they are based on
second-order perturbation theory.

After introducing our system-bath model in Section \ref{sec:model}, we
present in Sections \ref{sec:masterequation} and
\ref{secExactSolution} a master equation for weak system-bath coupling
and the exact time-evolution of the reduced system density operator,
respectively.  In Section \ref{sec:discussion}, we argue why the exact
solution complies with the master equation.

\section{System-bath model}
\label{sec:model}

We model the dissipative quantum system by coupling the central system to a
quantum heat bath that consists of harmonic oscillators, so that the total
system-bath Hamiltonian reads \cite{Leggett1987a,Hanggi1990a}
\begin{equation} \label{eqnH0}
H = H_\text{s} + H_\text{sb} + H_\text{b} \,,
\end{equation}
and $H_0 = H_\text{s} +H_\text{b}$ describes the system and the bath in the
absence of the coupling. While we will not specify the system Hamiltonian, we
employ for the bath an ensemble of independent harmonic oscillators with the
Hamiltonian
\begin{equation} \label{eqnHb}
H_\text{b} = \sum_k \hbar \omega_k b^\dagger_k b_k.
\end{equation}
Here, $b^\dagger_k$ and $b_k$ are the usual creation and annihilation operators
which obey the commutation relation $[b_k, b^\dagger_{k'}] = \delta_{kk'}$. We
assume that the system couples linearly to the bath with a hermitian system
operator $X$, so that the interaction Hamiltonian reads
\begin{equation} \label{eqnHsb}
H_\text{sb} = \hbar X \xi,
\end{equation}
with the effective bath coordinate
\begin{equation} \label{eqnBathCoordinate}
\xi = \sum_k \big( g_k b_k + g_k^\ast b^\dagger_k\big)\,.
\end{equation}
If the coupling operator $X$ commutes with the system Hamiltonian,
$[X,H_\text{s}]=0$, the coupling \eqref{eqnHsb} does not induce transitions
between system eigenstates and, thus, constitutes pure phase noise.  Henceforth
we shall focus on this case.

We choose an initial condition of the Feynman-Vernon type, i.e.~one
for which the total density operator $R$ at time $t=0$ can be factorised
into a system and a bath contribution $\rho$ and $\rho_\text{b}$,
respectively, i.e.
\begin{equation} \label{eqnInitialCondition}
R(0) = \rho(0) \rho_\text{b}(0).
\end{equation}
The bath itself is frequently assumed to be initially at thermal equilibrium.
However, if the initial expectation value of the coupling operator $X$ does not
vanish, the coupling \eqref{eqnHsb} entails a force on the bath oscillators.
Then the natural initial state $\rho_\text{b} = \rho_\text{b}(0)$ of the bath is
rather a displaced thermal state which falls in the class of non-squeezed
Gaussian states.  A convenient basis for these states is provided by the
coherent states $\{|\beta_k\rangle\}$ defined by the eigenvalue equation $b_k
|\beta_k\rangle = \beta_k |\beta_k \rangle$.  Owing to the overcompleteness of
this basis, any hermitian operator can be written in a diagonal form, which
assigns to each operator a $P$-function \cite{Perelomov1986a, Gardiner1991a}. In
particular, the bath density operator can be written as
\begin{equation} \label{eqnInitialBathState}
\rho_\text{b} = \int |\beta_1,\beta_2,\ldots\rangle\langle\beta_1\beta_2,\ldots|
\prod_k P_k(\beta_k,\beta_k^\ast)  \diff^2\beta_k
\,,
\end{equation}
where $\d^2\beta_k$ denotes integration over the complex plane. Henceforth, we
assume that the $P$-function of each oscillator $k$ is a Gauss function, such
that
\begin{equation} \label{eqnPfunction}
P_k(\beta_k,\beta_k^\ast) = \frac{1}{\pi n_k} \exp\left( \frac{-(\beta_k -
\bar\beta_k)(\beta_k-\bar\beta_k)^\ast}{n_k}\right)\,.
\end{equation}
As a central property of a Gaussian
state, all expectation values are fully determined by $\bar\beta_k =
\langle b_k\rangle_\text{b}$ and $n_k = \langle b_k^\dagger
b_k\rangle_\text{b} -|\langle b_k\rangle_\text{b}|^2$, where
\begin{equation} \label{eqnTraceBath}
\langle\ldots\rangle_\text{b} = \trace_\text{b}(\rho_\text{b} \ldots)
\end{equation}
denotes the expectation
value with respect to the bath state $\rho_\text{b}$.
An important particular case is the canonical ensemble of the bath at
temperature $T$ such that $\rho_\text{b}\propto
\exp(-H_\text{b}/k_\text{B}T)$, which corresponds to $\langle
b_k\rangle_\text{b}=0$ and $2n_k =
\coth(\hbar\omega_k/2k_\text{B}T)-1$.

The dynamics of the system plus the bath is governed by the Liouville-von
Neumann equation
\begin{equation} \label{eqnLiouville}
\dot{\Rt}(t) = \Lt \Rt(t)\,,
\end{equation}
with the Liouvillian $\tilde{\mathcal{L}}(\ldots) = - \im [ \widetilde
H_\text{sb}(t), \ldots ]/\hbar$. The tilde denotes the interaction-picture
representation with respect to $H_0$, i.e.\ $\tilde A(t) = U_0^\dagger(t) A
U_0(t)$, with the free propagator $U_0(t) = \exp(- \im H_0 t/\hbar)$. The
interaction-picture representation of the effective bath coordinate $\xi$ is
easily obtained from $\tilde b_k(t) = b_k\exp(-\im\omega_k t)$, while for pure
phase noise, $\widetilde X(t) = X$, owing to the relation $[X,H_\text{s}]=0$. For
the same reason, $H_\text{s}$ and $X$ possess a complete set of common
eigenstates $\{|n\rangle\}$, for which the respective eigenvalues are denoted by
$E_n$ and $X_n$.

We are exclusively interested in the state of the system, so our goal is to find
the reduced density operator $\rhot(t) = \trace_\text{b} \widetilde R(t)$.
In the subsequent sections, we derive explicit expressions for the reduced
dynamics.


\section{Master equation approach}
\label{sec:masterequation}

A common and successful approach to dissipative quantum dynamics is provided by
master equations, i.e.~differential or integro-differential equations of motion
for the reduced density operator
\cite{Gardiner1991a,vanKampen2001a,Breuer2002a}. There exist various formally
exact quantum master equations in time-convolutionless \cite{Fulinski1967a,Fulinski1968a,
Tokuyama1976a, vanKampen1974a, Hanggi1977a, Grabert1977a, Shibata1977a,
Shibata1978a, Hanggi1978, Hanggi1982a, Zerbe1995a} and time-nonlocal form
\cite{Nakajima1958a, Zwanzig1960a, Haake1973a} which, however, generally cannot
be solved explicitly and, thus, one often has to resort to a perturbative
treatment. The need for approximations can stem from the fact that the model
itself cannot be solved exactly, or from a technically difficult structure of
the master equation, or even both.

Here, we employ a time-convolutionless quantum master equation of the form
\begin{equation} \label{eqnTCLME}
\dot{\rhot}(t) = \mathcal{K}(t) \rhot(t)\,,
\end{equation}
with a time-dependent superoperator $\mathcal{K}(t)$. Note that there arises no
inhomogeneity since we are assuming a factorising initial state of the form
\eqref{eqnInitialCondition}, which leads to a linear equation of motion \cite{FonsecaRomero2004b}. Equation \eqref{eqnTCLME} is formally exact and
possesses an apparently simple form, but it generally cannot be solved
analytically.  Thus, it is convenient to expand the generator $\mathcal{K}(t)$
in powers of the interaction, i.e. $\mathcal{K}(t) = \sum_\ell
\mathcal{K}_\ell(t)$. By a direct calculation \cite{Kubo1962a,vanKampen1974b} or
by using a time convolutionless projection operator technique
\cite{Shibata1977a,Chaturvedi1979a,Breuer2001a} it is possible to obtain an
expression for the $\ell$th order generator $\mathcal{K}_\ell$. In doing so, we
assign to a superoperator $\mathcal{G}$ of the total system a reduced
superoperator
$\langle\mathcal{G}\rangle$ defined by its action on a system operator $Y$,
that is $\langle \mathcal{G} \rangle Y = \trace_\text{b} \{
\mathcal{G}(Y \rho_\text{b})\}$. With this notation, the time-convolutionless
generators read $\mathcal{K}_1(t)= \langle \Lt(t) \rangle$, and
\begin{equation} \label{eqnTCLGeneratorNthOrder}
\begin{split}
\mathcal{K}_\ell(t) &= \int_0^t \diff t_1 \int_0^{ t_1} \diff  t_2 \ldots
\int_0^{ t_{\ell-2}} \diff t_{\ell-1} \\ 
&\qquad \times \langle\langle \Lt(t) \Lt( t_1) \ldots
\mathcal{L}( t_{\ell-1})\rangle\rangle_\text{oc}\,,
\end{split}
\end{equation}
for $\ell=2,3,\ldots$
The symbol $\langle\langle \ldots \rangle\rangle_\text{oc}$ denotes an
ordered cumulant \cite{vanKampen1974b,vanKampen2001a}, i.e.~a sum of 
certain products of reduced superoperators of the form
$\langle \Lt(t) \Lt(t_1) \ldots \rangle$.

The fact that the Liouvillians at different times generally do not
commute makes it practically impossible to write down an explicit
expression for the $\ell$th cumulant for large $\ell$. However, for
weak system-bath coupling it is possible to neglect higher than second
order terms in the expansion of the generator, i.e.~we may approximate
$\mathcal{K}(t) \approx \mathcal{K}_1(t) + \mathcal{K}_2(t)$.
Fortunately the second time-ordered cumulant takes the simple form
$\langle\langle \Lt(t) \Lt( t_1) \rangle\rangle_\text{oc} = \langle \Lt(t) \Lt(
t_1) \rangle - \langle \Lt(t) \rangle \langle \Lt( t_1) \rangle$.
Considering now explicitly the interaction Hamiltonian \eqref{eqnHsb}, we obtain
the ``standard'' time-local weak-coupling equation
\begin{equation} \label{eqnME}
\begin{split}
\dot{\rhot} (t)
={}& 
-\im \langle \tilde\xi (t)\rangle_\text{b} [X,\rhot(t)] \\
& -\int_0^t\diff\tau\, \Big( \mathcal{S}(t,t-\tau)
  [X,[X,\rhot(t)]]
\\
& \qquad \quad \quad
+ \im \mathcal{A}(t,t-\tau) [X,\{X,\rhot(t)\}]
\Big) \,,
\end{split}
\end{equation}
where $\{A,B\} = AB+BA$ denotes the anti-commutator, and where we have defined 
the symmetric and anti-symmetric correlation functions
\begin{align}
\label{sym}
\mathcal{S}(t,t')
&= \frac{1}{2} \langle \{\Delta\tilde\xi(t),\Delta\tilde\xi(t')\}
   \rangle_\text{b}, \\ \label{anti}
\mathcal{A}(t,t')
&= \frac{1}{2} \langle [\Delta\tilde\xi(t),\Delta\tilde\xi(t')]
   \rangle_\text{b},
\end{align}
of the operator-valued fluctuation $\Delta\tilde\xi(t) = \xi(t)-
\langle\xi(t)\rangle_\text{b}$.  Note that for a Gaussian initial state
$\rho_\text{b}$ of the bath, the mean value $\langle \xi(t) \rangle_\text{b}$ of
the bath coordinate in general does not vanish.  Thus, it explicitly appears in
the master equation \eqref{eqnME}.

It is convenient to expand the master equation \eqref{eqnME} into the eigenbasis
of the system-bath interaction.  We then obtain for a matrix element $\langle m |
\rhot | n\rangle = \rhot_{mn}$ the differential equation
\begin{equation} \label{eqnMEMatrixElements}
\begin{split}
\dot{\rhot}_{mn}(t)
={}& \Big [ -\im  (X_m - X_n)
   \langle \tilde\xi(t) \rangle_\text{b}  \\
& -(X_m-X_n)^2  \int_0^t\diff\tau\,
  \mathcal{S}(t,t-\tau) \\
& - \im (X_m^2 - X_n^2)
\int_0^t\diff\tau\,\mathcal{A}(t,t-\tau) \Big]
\rhot_{mn} (t)\,.
\end{split}
\end{equation}
For the diagonal matrix elements $\rhot_{nn}$ the right-hand side of
Eq.~\eqref{eqnMEMatrixElements} vanishes, i.e.~the populations remain constant
in time as one expects for a pure dephasing model. This implies that generally
the system  will not reach thermal equilibrium.  Although no energy is exchanged
and, thus, the interaction is dissipationless, the relative phases between
eigenstates will be randomised. As a consequence, off-diagonal elements of the
reduced density matrix may decay, which reflects the process of decoherence. 

Using the coherent state representation \eqref{eqnPfunction} for a Gaussian bath
state, we can evaluate the mean value of the bath coordinate $\tilde\xi$ and the
correlation functions in an explicit form and find
\begin{align}
\langle \tilde\xi(t) \rangle_\text{b} 
&= \sum_k \left( g_k\bar\beta_k\e^{-\im \omega_k t} + g_k^\ast
  \bar\beta_k^\ast \e^{\im \omega_k t} \right)
\\ \label{eqn2ndOrderRe}
\mathcal{S}(t,t-\tau)
&=  \sum _k | g_k |^2 \cos (\omega_k \tau) (1 + 2 n_k)
\\ \label{eqn2ndOrderIm}
\mathcal{A}(t,t-\tau) 
&= - \sum_k | g_k |^2 \sin (\omega_k \tau)\,.
\end{align}
Note that the correlation functions $\mathcal{S}(t,t')$ and
$\mathcal{A}(t,t')$ depend on time differences $t-t'$, only, with
$\mathcal{A}(t,t')$ independent of the initial state of the bath. If the
correlation functions vanish sufficiently fast, it is possible to
employ a Markov approximation, i.e.~to extend the upper integration
limits in Eq.~\eqref{eqnMEMatrixElements} to infinity. Then, the
master equation~\eqref{eqnMEMatrixElements} reduces to the standard
Bloch-Redfield equation \cite{Blum1996a} and, moreover, is of
Lindblad form, so that the complete positivity of
the reduced density operator is conserved even for arbitrarily short
times  \cite{Kraus1970, Lindblad1976a}.


\section{Exact solution} \label{secExactSolution}

The dynamics of a system subject to pure phase noise can, at least in principle,
be solved analytically  \cite{Palma1996a, Luczka1990a, vanKampen1995a,
Duan1998a, Mozyrsky1998a, Braun2001a, Krummheuer2002a, Doll2006a, Doll2007b,
Solenov2007a}. The formal solution of the Liouville-von Neumann equation
\eqref{eqnLiouville} after tracing over the bath's degrees of freedom reads
\begin{equation} \label{eqnReducedDensityMatrixExact}
\rhot(t) = \trace_\text{b} \{ U(t) R(0) U^\dagger(t) \} ,
\end{equation}
with the propagator
\begin{equation} \label{eqnPropagator}
U(t) = \mathcal{T} \exp \biggl\{
\frac{1}{\im\hbar} \int_0^t\diff\tau\, \widetilde H_\text{sb}(\tau)\bigg\}
\end{equation}
and the time-ordering operator $\mathcal{T}$.  Although the consideration of the
time ordering can often be quite cumbersome, it nevertheless can be accomplished
for the model discussed here. We deferred the explicit derivation to Appendix
\ref{secDerivationOfExactSolution}, where we find for the elements of the
reduced density matrix the exact expression
\begin{equation} \label{eqnReducedDensityMatrixElements}
\begin{split}
\rhot_{mn}(t) &= \rhot_{mn}(0)\e^{\im  (X_m^2 - X_n^2) 
\phi(t)- (X_m-X_n)^2\sum_k |z_k(t)|^2/2} \\
& \quad \times \prod_k \chi_k\bigl\{z_k(t) [X_m-X_n], z_k^\ast(t)
[X_m-X_n]\bigr\}\,,
\end{split}
\end{equation}
with the time-dependent phase
\begin{equation} \label{eqnExactPhase1}
\phi(t) =  \sum_k \frac{|g_k|^2}{\omega_k^2}\left[ \omega_k t - \sin(\omega_k
t)\right]\,,
\end{equation}
the quantum characteristic function \cite{Gardiner1991a}
\begin{equation} \label{eqnCharacteristicFuntion}
\chi_k(\lambda_k,\lambda_k^\ast) = \int \e^{\lambda_k \beta_k^\ast - \lambda_k^\ast \beta_k} P_k(\beta_k,\beta_k^\ast) 
\diff^2\beta_k\,,
\end{equation}
and the time-dependent complex number $z_k(t) = g_k^\ast
[1-\exp(\im\omega_kt)]/\omega_k$.  To arrive at
Eq.~\eqref{eqnReducedDensityMatrixElements}, we only used the assumption that
initially the system and the bath are uncorrelated and that the initial bath
state $\rho_\text{b}$ factorises with respect to the
modes $k$ [see Eqs.~\eqref{eqnInitialCondition} and
\eqref{eqnInitialBathState}]. For the case of a Gaussian distribution
of the bath modes, it is possible to calculate the integral
in the characteristic function \eqref{eqnCharacteristicFuntion} explicitly. We
finally find the exact time evolution of the reduced matrix element
\begin{equation} \label{eqnExactSolution}
\begin{split}
\rhot_{mn}(t) &= \rho_{mn}(0) 
\exp\Bigl\{ - (X_m -X_n)^2 \Lambda(t) \\ 
&\quad  + \im \bigl[(X_m^2 - X_n^2)\phi(t) 
+(X_m - X_n) \varphi(t)\bigr]\Bigr\}\,,
\end{split}
\end{equation}
with the phases $\phi(t)$ defined in Eq.~\eqref{eqnExactPhase1} and
\begin{equation} \label{eqnExactPhase2}
\varphi(t) = 2\sum_k \Imag \Big(
\frac{\bar\beta_k^\ast g_k^\ast}{\omega_k} \bigl[ 1- \e^{\im \omega_k t} \bigr]
\Big)\,.
\end{equation}
The time dependent damping amplitude $\Lambda(t)$ does not depend on the mean
values $\bar\beta_k$ of the bath modes and reads
\begin{equation} \label{eqnExactDampingAmplitude}
\Lambda(t) = \sum_k |g_k|^2 \frac{1-\cos(\omega_k t)}{\omega_k^2}
(1+2n_k)\,.
\end{equation}
Note that a similar result was obtained recently for a bath initially
in a squeezed thermal state \cite{Banerjee2007a}. 

Upon computing the time-derivative of the exact
solution~\eqref{eqnExactSolution} and noting that the relations
\begin{align} \label{relation1}
\dot \varphi (t)
&= \langle \xi (t) \rangle_\text{b}
\\ \label{relation2}
\dot \phi(t)
&=
- \int_0^t\diff \tau\,\mathcal{A}(t,t-\tau)
\\  \label{relation3}
\dot \Lambda (t) &= \int_0^t\diff \tau \, \mathcal{S}(t,t-\tau)
\end{align}
hold, we find the surprising fact that the exact solution obeys the quantum
master equation \eqref{eqnMEMatrixElements}! Or put differently, for pure
phase noise, the exact result can be obtained within second-order perturbation
theory from the master equation \eqref{eqnME}. For large times $t$, this master
equation becomes the standard Markovian Bloch-Redfield equation. Thus, we find
that the latter contains the exact long-time limit of the rates
\eqref{relation1}--\eqref{relation3}.

Before discussing the relation between both approaches in more
detail, we like to close this section by writing the exact solution
\eqref{eqnExactSolution} also in terms of the usual bath spectral
density \cite{Leggett1987a, Hanggi1990a}
\begin{equation}
J(\omega) = \sum_k |g_k|^2 \delta(\omega-\omega_k)\,.
\end{equation}
For the important special case of a heat bath that is initially at thermal
equilibrium, we find $\varphi(t)=0$, while the phase $\phi(t)$ and the damping
amplitude $\Lambda(t)$ read
\begin{align}
\phi(t) &= \int \diff \omega  \, J(\omega) \frac{\omega t - \sin(\omega
t) }{\omega^2}\,
\\
\Lambda(t) &= \int \diff \omega  \, J(\omega) \frac{1-\cos(\omega t)}{\omega^2}
\coth\Big(\frac{\hbar\omega}{2k_\text{B}T}\Big)\,.
\end{align}


\section{When second order is exact}
\label{sec:discussion}
We have seen that the time-local master equation \eqref{eqnTCLME}
derived within second order perturbation theory provides the exact
time evolution of the reduced density matrix, which implies that in
the expansion of the Liouvillian $\mathcal {K}(t)$, all higher
order contributions vanish. This on the one hand nicely simplifies
practical calculations. On the other hand, it poses the question
whether we face a coincidence or whether there is any
profound reason for the equivalence.
In order to underline the latter point of view, we now demonstrate that for
phase noise, the time-ordered cumulant in the $\ell$th order generator
\eqref{eqnTCLGeneratorNthOrder} is proportional to the usual classical cumulant
of the initial bath state. Consequently, we can argue that for the Gaussian
initial state \eqref{eqnPfunction}, the series $\mathcal{K}(t) =
\sum_\ell\mathcal{K}_\ell(t)$ terminates after $\ell=2$, which implies that the
second-order time-local master equation \eqref{eqnME} is exact.

We start out by defining averages with respect to the $P$-function as
\begin{equation} \label{eqnAverageOverPFunction}
\langle \cdots \rangle_P
= \int\cdots \prod_k  P(\beta_k,\beta_k^\ast)\diff^2\beta_k\,.
\end{equation}
With this notation, the exact solution~\eqref{eqnReducedDensityMatrixElements}
reads
\begin{equation} \label{eqnReducedDensityMatrixElements2}
\rhot_{mn}(t) = \biggl\langle \exp\biggl\{\int_0^t\diff\tau\,
f_{mn}(\tau)\biggr\}
\biggr\rangle_P\,\rhot_{mn}(0)\,,
\end{equation}
with the complex valued function
\begin{equation}
\begin{split}
f_{mn}(t) &= \im (X_m^2 - X_n^2)\phi(t) -
     \sum_k \Big\{ (X_m - X_n)^2 \frac{|z_k(t)|^2}{2}\\
     &\qquad - (X_m - X_n) [z_k(t) \beta_k^\ast -
     z_k^\ast(t) \beta_k] \Big\} .
\end{split}
\end{equation}
The average in Eq.~\eqref{eqnReducedDensityMatrixElements2} is obtained from a
distribution function for the c-numbers $\beta_k$. Thus, it can be formally
considered as the averaged solution of a stochastic differential equation that
obeys a time-local differential equation of the form~\eqref{eqnTCLME}, but with
the generator $\mathcal{K}$ now being a c-number, not an operator.
Thus, we can adapt the line of argumentation given by van
Kampen  for classical Gaussian stochastic processes \cite{vanKampen1974b}:
Differentiating the Taylor expansion of
Eq.~\eqref{eqnReducedDensityMatrixElements2}, we find
\begin{equation}
\begin{split}
&\dot{\rhot}_{mn}(t)
= \biggl[ \langle f_{mn}(t) \rangle_P +
  \int_0^t \!\diff t_1\langle f_{mn}(t) f_{mn}(t_1)\rangle_P \\
&+ \int_0^t \!\diff t_1\! \int_0^{ t_1}\! \diff t_2\,\langle
f_{mn}(t) f_{mn}( t_1)f_{mn}( t_2) \rangle_P 
+ \ldots \, \biggr] \rhot_{mn}(0).
\end{split}
\end{equation}
A time-local equation of motion for $\rhot_{mn}(t)$ can be obtained by
inserting $\rhot_{mn}(0)$ from
Eq.~\eqref{eqnReducedDensityMatrixElements2}, which yields \cite{vanKampen1974b}
\begin{equation} \label{eqnTCLMEMatrixElements}
\begin{split}
\dot{\rhot}_{mn}(t)
={}& \sum_{\ell=1}^\infty \int_0^t \diff t_1 \int_0^{ t_1}
  \diff t_2 \ldots \int_0^{ t_{\ell-2}} \diff t_{\ell-1} \\
& \times \langle\langle f_{mn}(t) f_{mn}(t_1) \ldots
  f_{mn}( t_{\ell-1})\rangle\rangle_P\, \rhot_{mn}(t)\,,
\end{split}
\end{equation}
where $\langle\langle\ldots\rangle\rangle_P$ denotes the cumulants
with respect to the $P$-function.  Note that for the cumulants of a
classical process, time-ordering is not relevant
\cite{vanKampen1974b}.  Thus the only difference of this expansion and
the one in Eq.~\eqref{eqnTCLGeneratorNthOrder} for the quantum master
equation is that the latter contains time-ordered cumulants.

For a Gaussian $P$-function, all cumulants of $\beta_k$ and $\beta_k^\ast$
beyond second order vanish \cite{Risken1984a}.  Since $f_{mn}(t)$ is linear in
these variables, the same is true for the cumulants in
Eq.~\eqref{eqnTCLMEMatrixElements} and, consequently, only the terms with
$\ell=1,2$ contribute to this expansion.  Evaluating the expansion coefficients
explicitly, one finds that they are identical to those of the second-order
time-local master equation \eqref{eqnMEMatrixElements}.

The equivalence of the second-order
master equation and the exact solution is based on two requirements:
First, the coupling operator $X$ needs to be diagonal in the
eigenbasis of the system, so that its interaction-picture
representation is time-independent, $\widetilde X(t)=X$, and,
thus, it can be effectively treated as a c-number.  Hence the
quantum mechanical time-ordering affects only the bath coordinate
$\tilde\xi(t)$ for which we can express multi-time expectation values
as cumulants of the $P$-function. In that way, we can circumvent the tedious
task of normal-ordering the operators $b_k$ and $b_k^\dagger$.  With
this precondition, secondly, the Gaussian initial state of the bath
ensures that the cumulant expansion terminates after the second order
and agrees with the expansion of the master equation \eqref{eqnTCLME}.
For any non-Gaussian state, infinitely many higher-order cumulants are
non-zero both in the classical case \cite{Marcinkiewicz1938a,
Pawula1967a, Hanggi1980a} and in the quantum mechanical case
\cite{Rajagopal1974a,Titulaer1975a,Schack1990a}.  Consequently, the expansion of the Liouvillian
is of infinite order and any truncation represents an approximation.

Let us finally stress that the second-order Nakajima-Zwanzig master
equation \cite{Nakajima1958a, Zwanzig1960a}, which was not considered
in this work, can be expressed in terms of cumulants (the so-called
partial cumulants), as well \cite{Yoon1975a,Shibata1980a}. Note that
the ordered and the partial cumulants up to second order
coincide. However, the second-order Nakajima-Zwanzig equation is not
of a time local form and therefore cannot yield the exact result for
the model discussed here \cite{Solenov2004}.  Thus, phase noise
constitutes an example for which the time-local approach is more
accurate than the time-nonlocal one when comparing their perturbation
expansions up to the same order. This outcome is in agreement with some
recent findings for harmonic oscillator baths \cite{Yan2005a,Royer2003a} and 
for spin baths \cite{Breuer2004,Krovi2007}. For example for a
two-level system coupled via XY Heisenberg interaction to a spin bath, the
differences of both approaches have been analysed quantitatively \cite{Breuer2004}.
Nevertheless, we do not give a general recommendation in favour of one or the other
approach because the quality of each seems to be model dependent
\cite{Yoon1975a,Mukamel1978a,Reichman1997a,Schroeder2007a}.

\section{Conclusion}

Quantum systems under the influence of pure phase noise represent an important
special case of dissipative quantum mechanics owing to the existence of an exact
solution.  Moreover, on short time scales, on which the coherent system dynamics
cannot manifest itself, the behaviour of the phase noise model is even generic
\cite{Braun2001a}.  Here, we have presented the explicit exact solution for a
quantum system under the influence of phase noise with a general Gaussian
initial bath state. Thereby, we have demonstrated that the coherence decay is
determined by the symmetric bath correlation function, while the anti-symmetric
correlation function gives rise to a time-dependent phase shift.
In turn, from the exact relations
\eqref{relation1}--\eqref{relation3}, one can obtain information
about the spectral properties of the bath by comparing our results
with the experimentally observed dephasing at short times.

Despite the exact solvability of the phase noise problem, it is often
convenient to study the resulting dephasing within a master equation
approach based on second-order perturbation theory in the system-bath
coupling.  
For the (time-nonlocal) Nakajima-Zwanzig equation, this constitutes
an approximation. 
For the time-local version of such a master equation, by contrast, we
have found that it provides the exact solution.
After noticing that this facilitates practical calculations, one might
wonder why and when this equivalence holds true. By mapping the
initial bath density operator to a $P$-function, we showed that a
formal expansion of the time-local master equation for phase noise is
in fact an ordinary cumulant expansion.  Consequently, for a Gaussian
initial bath state, all terms beyond the second order vanish, so that
the master equation becomes exact.
Thus for a bosonic heat bath, there are two conditions for
the exact agreement: First, the
system-bath coupling must commute with the system Hamiltonian
constituting the case of pure phase noise and, second, the initial
state of the bath must correspond to a Gaussian $P$-function. If one
of these conditions is violated, there might still exist an exact
solution, but it can no longer be obtained within second-order
perturbation theory, as for example is the case for the dissipative
harmonic oscillator \cite{Grabert1984b, Grabert1988a}.

The second-order time-local master equation employed in this work agrees with
the exact solution at any time. 
In particular in the long-time limit, it
becomes Markovian and identical to the standard Bloch-Redfield master equation,
which for pure phase noise is of Lindblad form.  This also explains the
previously observed ``excellent agreement'' \cite{Lidar2001a} between the exact
dynamics and results obtained within Bloch-Redfield theory.

\section*{Acknowledgements}
This work was supported by Deutsche Forschungsgemeinschaft through
SFB~484 and SFB~631. PH and SK acknowledge funding by the DFG
excellence cluster ``Nanosystems Initiative Munich''.

\appendix

\section{Derivation of the exact solution} \label{secDerivationOfExactSolution}

In this appendix, we outline the derivation of the exact reduced
dynamics discussed in Sec.~\ref{secExactSolution}.
As a first step, we perform a transformation to the interaction
picture with respect to $H_0$, so that the coupling Hamiltonian
\eqref{eqnHsb} becomes
\begin{equation}
\widetilde H_\text{sb}(t) = \widetilde V(t) + \widetilde V^\dagger(t)\,,
\end{equation}
with $\widetilde V(t) = \hbar X \sum_k g_k b_k \exp(-\im\omega_k t)$ and
the hermitian system operator $X$. The two operators 
$\widetilde V$ and $\widetilde V^\dagger$ do not commute. However, their
commutator is an operator in the Hilbert space of the system, while
being a c-number in the bath Hilbert space.  To be specific, one
obtains $[\widetilde V(t), \widetilde V^\dagger(t')] = f(t-t')$, where
\begin{equation}
f(t) = \hbar^2 X^2 \sum_k |g_k|^2 \e^{-\im \omega_k t}\,.
\end{equation}
Hence, we can use the Baker-Campbell-Hausdorff formula (see
Ref.~\cite{Gardiner1991a}) to express the time-ordered propagator
\eqref{eqnPropagator} as a product of two exponentials
\begin{equation} \label{eqnPropagatorSplitted}
\begin{split}
U(t) &= \exp\biggl\{ \frac1{\im\hbar} \int_0^t\diff\tau\, \widetilde
H_\text{sb}(\tau) \biggr\} \\
&\quad \times \exp\biggl\{-\frac{1}{\hbar^2}\int_0^t\diff\tau \int_0^\tau
\diff\tau'\,f(\tau-\tau')\\
&\qquad\qquad \times [\theta(\tau-\tau') - \theta(\tau'-\tau)] \biggr\}\,.
\end{split}
\end{equation}
The first exponential in Eq.~\eqref{eqnPropagatorSplitted} can be written as
\begin{equation}
\exp\biggl\{ \frac1{\im\hbar} \int_0^t\diff\tau\, \widetilde
H_\text{sb}(\tau) \biggr\} = \prod_k D_k(z_k X)
\end{equation}
with the complex number $z_k = g_k^\ast [1-\exp(\im\omega_k t)] /
\omega_k$ and the displacement operators $D_k(Y) = \exp(Y b_k^\dagger
- Y^\dagger b_k)$. The second exponential in
Eq.~\eqref{eqnPropagatorSplitted} provides the time-dependent phase
factor $\exp\{\im X^2 \phi(t)\}$ with
\begin{equation}
\phi(t) = \int_0^t\diff\tau\, \int_0^{\tau}\diff \tau'\, \sum_k |g_k|^2
\sin(\omega_k \tau')\,.
\end{equation}

We will now consider the elements of the reduced density matrix in the
eigenbasis $\{|n\rangle\}$ of the system Hamiltonian. Using the initial
condition \eqref{eqnInitialCondition}, Eq.~\eqref{eqnReducedDensityMatrixExact}
becomes
\begin{align}
\rhot_{mn}(t) ={}& \trace_\text{b}\langle m | U(t) \rho(0)\rho_\text{b}
U^\dagger(t) | n \rangle \\
={}& \rho_{mn}(0) \e^{\im [X_m^2 - X_n^2] \phi(t)}\nonumber \\
   & \times \biggl\langle
     \prod_k D_k^\dagger(X_n z_k) \prod_{k'} D_{k'}^\dagger(X_m
     z_{k'})\biggr\rangle_\text{b}\,,
\end{align}
where $\langle\ldots\rangle_\text{b} = \trace_\text{b}(\rho_\text{b}\ldots)$.
By virtue of the relations $D_k^\dagger(Y) = D_k^\dagger(-Y)$ and
\begin{equation}
D_k(Y) D_k(Z) = D_k(Y+Z) \exp\{(Y Z^\dagger - Y^\dagger Z)/2\}\,,
\end{equation}
which hold for any commuting system operators $Y$ and $Z$, we obtain
\begin{equation} \label{eqnReducedDensityMatrixElementsGeneral}
\begin{split}
\rhot_{mn}(t)
={}& \rho_{mn}(0) \e^{\im [X_m^2 - X_n^2] \phi(t)  + \im \eta_{mn}}\\
&\times \biggl\langle \prod_k D_k(z_k[X_m - X_n]) \biggr\rangle_\text{b}\,.
\end{split}
\end{equation}
Note that an additional phase $\eta_{mn}$ $=$ $2\sum_k |z_k|^2
\Imag(X_m^\ast X_n)$ vanishes, since the system operator $X$ is hermitian.

It remains to evaluate the expectation value in the second line of
Eq.~\eqref{eqnReducedDensityMatrixElementsGeneral}. This is readily
established by writing the bath state $\rho_\text{b}$ in its
$P$-function representation [see Eq.~\eqref{eqnInitialBathState}] and
noticing that expectation values of normal ordered products of
annihilation and creation operators are identical to the moments of
the $P$-function, where $b^\dagger_k$ and $b_k$ have to be replaced by
$\beta_k^*$ and $\beta_k$, respectively \cite{Gardiner1991a}. Thus, by
using the Baker-Campbell-Hausdorff formula
\begin{equation}
\exp(\lambda b_k^\dagger - \lambda_k^\ast b_k) = \exp(\lambda b_k^\dagger) \exp(\lambda^\ast
b_k) \exp( - |\lambda|^2/2)
\end{equation}
for each mode $k$, we write the second line of
Eq.~\eqref{eqnReducedDensityMatrixElementsGeneral} in its normal-ordered form
\begin{equation} \label{eqnNormalOrderedDisplacementOperators}
\e^{ - [X_m - X_n]^2 \sum_k |z_k|^2/2} \biggl\langle \prod_k \e^{z_k [S_m-S_n]
b_k^\dagger} \e^{z_k^\ast [S_m-S_n] b_k}
\biggr\rangle_P\,,
\end{equation}
where $\langle \ldots \rangle_P$ denotes the average
\eqref{eqnAverageOverPFunction} with respect to the $P$-function.
Using also the fact that the $P$-function \eqref{eqnPfunction}
factorises with respect to the modes $k$, we arrive at 
Eq.~\eqref{eqnReducedDensityMatrixElements}.


\end{document}